\documentclass[3p,times,procedia]{elsarticle}
\usepackage{ecrc}
\usepackage{wrapfig}
\usepackage{graphicx}
\usepackage[bookmarks=false]{hyperref}
    \hypersetup{colorlinks,
      linkcolor=blue,
      citecolor=blue,
      urlcolor=blue}

\volume{00}

\firstpage{1}

\journalname{Procedia Computer Science}

\runauth{Antonio Fernández, Suzan Awinat}

\jid{procs}

\usepackage{amssymb}

\usepackage[figuresright]{rotating}

\begin{document}
\begin{frontmatter}



\dochead{The 15th International Conference on Emerging Ubiquitous Systems and Pervasive Networks \\(EUSPN 2024) \\ October 28-30, 2024, Leuven, Belgium}

\title{Multimodal Sentiment Analysis based on Video and Audio Inputs}


\author[a]{Antonio Fernández} 
\author[a]{Suzan Awinat}
\corref{cor1}

\address[a]{Cyphy Life, Robotics \& AI Lab, School of science and technology, IE University, Madrid, Spain}

\begin{abstract}

Despite the abundance of current researches working on the sentiment analysis from videos and audios, finding the best model that gives the highest accuracy rate is still considered a challenge for researchers in this field.

The main objective of this paper is to prove the usability of emotion recognition models that take video and audio inputs. The datasets used to train the models are the CREMA-D dataset for audio and the RAVDESS dataset for video. The fine-tuned models that been used are: Facebook/wav2vec2-large for audio and the Google/vivit-b-16x2-kinetics400 for video. The avarage of the  probabilities for each emotion generated by the two previous models is utilized in the decision making framework. After disparity in the results, if one of the models gets much higher accuracy, another test framework is created. The methods used are the Weighted Average method, the Confidence Level Threshold method, the Dynamic Weighting Based on Confidence method, and the Rule-Based Logic method. This limited approach gives encouraging results that make future research into these methods viable.
\end{abstract}

\cortext[cor1]{Antonio Fernández}
\end{frontmatter}

\email{afernandez.ieu2020@student.ie.edu}



\section{Introduction}
\label{introduction}

The current landscape of sentiment analysis is quickly being improved, thanks to the addition of transformers. In the last two years, several papers have been published in this field and show how this new approach can lead to better results. But from all of the approaches that have been presented, our model will try to find the balance that leads to more accurate audio/video sentiment analysis.

To find out what has been tried already and what are some possible improvements that we can make to our approach, the table (Table 1) compares between some of the most recent sentiment analysis published models.

\begin{table}[h!]
    \centering
    \caption{Summary of Sentiment Analysis Models}
    \begin{tabular}{p{3cm} p{5cm} p{5cm} p{2cm}}
        \toprule
        Model & Description & Performance & Opinion\\
        \colrule
        TVLT: Textless Vision-Language Transformer [1] & Focuses solely on video and audio data. Reconstructs masked parts of video frames and audio spectrograms and aligns the two modalities through contrastive learning. & Performs comparably to text-based models on most emotions while outperforming them on some and being more computationally efficient. & Shows that our approach is viable. \\
        \colrule
        Enhanced Video Analytics for Sentiment Analysis Based on Fusing Textual, Auditory and Visual Information [2] & Sentiment analysis on social media tends to focus on text and is not suitable for complex languages like Arabic. Developed a multimodal approach that combines text, audio, and video features using the built dataset with Arabic videos and diverse speakers. & Performs better than unimodal approaches and the score based function generally yields better results than decision-based fusion. & Shows that combining different modalities can lead to better results. \\
        \colrule
        Fusing Audio, Textual, and Visual Features for Sentiment Analysis of News Videos [3] & Automates the analysis of news videos' sentiment through multimodal features. Combines audiovisual features from the video and text sentiment scores from closed captions to compute tension level. & 84\% accuracy in classifying tension levels and concludes that facial expressions were the most informative visual cue. & Shows that facial features tend to be the most important visual cue. \\
        \colrule
        Multimodal Video Sentiment Analysis Using Deep Learning Approaches, a Survey [4] & Comprehensive overview of deep learning models for video sentiment analysis. State-of-the-art models for video, audio and text were tested on the CMU-MOSI and CMU-MOSEI datasets. & Concludes that the best performance architecture was the Multi-Modal Multi-Utterance architecture and shows that the bimodal approach tended to lead to better results. & Shows that focusing on two modalities can lead to better results. \\
        \colrule
        TETFN: A Text Enhanced Transformer Fusion Network for Multimodal Sentiment Analysis [5] & Addresses the imbalance between modalities in multimodal sentiment analysis while enhancing the role of text information. Uses Vision Transformer for video, BERT for text, and COVERAP for audio. & TETFN outperforms state-of-the-art methods and facial visual features are more informative than general visual features. & Reinforces the importance of facial visual features. \\
        \colrule
        Transformer-based Feature Reconstruction Network for Robust Multimodal Sentiment Analysis [6] & Addresses the challenge of incomplete data in multimodal sentiment analysis and improves robustness. Uses feature extraction to process incomplete modality sequences and then does self-attention and linear transformation to fill in missing information. & Achieves strong performance even with missing modalities and the reconstruction module effectively captures semantics of missing features. & Shows that missing information can be handled while still having a strong model. \\
        \botrule
    \end{tabular}
\end{table}
In this paper, we aimed to find a balance between the approaches of these papers to get the best possible results. The models would leverage the advantages of the new transformer architecture for better prediction quality. As it was demonstrated, the first approach is focusing mainly on a bimodal architecture that takes into account video and audio inputs. In our approach, the videos for the training should showcase the facial expressions of the participants to get more precise predictions. Then, the architecture would need to include a scoring system to give the final output.
\section{ Methodologies and Preliminary Analyses}

\subsection{ Audio Classification Dataset}

The dataset selected for fine-tuning the audio classification task is the CREMA-D dataset [7]. This dataset contains 7442 voice clips from 91 actors, of which 48 are males and 43 are females. Furthermore, these actors are between the ages of 20 and 74 and come from a variety of races and ethnicities such as African American, Asian, Caucasian, Hispanic, and Unspecified.

The actors are instructed to say out loud a selection of 12 sentences. Each sentence is presented using one of the six available emotions: Anger, Disgust, Fear, Happy, Sad, and Neutral. Furthermore, each emotion could be represented by a different intensity level. These intensity levels are divided into Low, Medium, High, or Unspecified.

The labels about each audio clip can be found in the name of said clip. For example, the name of an audio clip looks something like this: "1001\_IEO\_ANG\_HI.wav". The first four numbers are an id that refers to the actor that dubbed the voice line. The next three letters refer to the sentence that was spoken by the actor. The next three capital letters refer to what emotion the audio clip represents. And the last two capital letters represent the intensity of the emotion. 

To standardize the labels of this dataset with the ones that were selected for the video, the labels were changed from ANG, DIS, FEA, HAP, NEU, and SAD to anger, disgust, fearful, happy, neutral, and sad respectively.

After performing some exploratory data analysis, it was discovered that most of the emotions provided by this dataset have around 1300 entries each. The only exception was the neutral emotion, which had around 1100 entries.

As for the intensity of the labels, it was discovered that 81.7\% were labeled as unspecified (XX). This is not a problem for the project since the overall intensity of the emotion is not a factor that is being taken into account.

In the end this dataset was chosen because of its variety of audio samples and the overlapping concordance of the emotion labels with the video dataset.

\subsection{ Video Classification Dataset}

The dataset selected for fine-tuning the video classification task was the RAVDESS dataset [8]. This dataset contains 7356 files with each being rated at least 10 times on emotional validity, intensity, and genuineness. The dataset itself contains 24 professional actors which are divided between 12 males and 12 females. Each of them vocalizes two lexically matched statements in a neutral North American accent. 

The dataset is divided into four folder categories. You have one audio speech folder and one audio song folder. Likewise, we also have one video speech folder and one video song folder. The difference between the song and speech categories is that the song categories have the actors sort of sing the sentences, while the speech categories have them say the sentences in normal speech. The speech categories for emotion are neutral, calm, happy, sad, angry, fearful, surprise, and disgust expressions. Each emotion can be presented at two levels of intensity (normal and strong).

And finally, the RAVDESS official website provides an example of how each filename is structured:

“Filename identifiers:
\begin{itemize}
    \item \textbf{Modality} (01 = full-AV, 02 = video-only, 03 = audio-only).
    \item \textbf{Vocal channel} (01 = speech, 02 = song).
    \item \textbf{Emotion} (01 = neutral, 02 = calm, 03 = happy, 04 = sad, 05 = angry, 06 = fearful, 07 = disgust, 08 = surprised).
    \item \textbf{Emotional intensity} (01 = normal, 02 = strong). NOTE: There is no strong intensity for the 'neutral' emotion.
    \item \textbf{Statement} (01 = "Kids are talking by the door", 02 = "Dogs are sitting by the door").
    \item \textbf{Repetition} (01 = 1st repetition, 02 = 2nd repetition).
    \item \textbf{Actor} (01 to 24. Odd-numbered actors are male, even-numbered actors are female)."
\end{itemize}

After some exploratory data analysis, we discovered that the emotions also appear to have a similar number of entries except for the neutral function which appears to have half the number of entries. It was also discovered that 53.8\% of the entries have a normal intensity while the remaining 46.2\% have a strong intensity. Lastly, and, that for each sentence the ratio of emotions tends to be similar.

It is also worth noting that the videos themselves are duplicated in the sense that each video with audio, has a variant without audio. To train our model we used only the videos that don’t have audio included since we are only focusing on the visual part.

To also have some data to test our final framework, the dataset was randomly sampled to get 105 files that have both video and audio. However, since the dataset is duplicated, the video-only counterparts from those 105 files were also removed from the training portion of the dataset.

In the end this dataset was chosen because of its variety of actors and the overlapping concordance of the emotion labels with the audio dataset.

\subsection{ Models and Framework}

Moving on, the models selected for fine-tuning into our specific task were the wav2vec2-large [9] from Facebook for the audio classification task and the vivit-b-16x2-kinetics400 [10] from Google for the video classification task.

The wav2vec2-large model is composed of a multi-layered convolutional feature encoder which receives an input of raw audio and outputs latent speech representations for a set amount of time steps.

The vivit-b-16x2-kinetics400 by Google is a transformer-based video classification model. It takes as an input a sample of video frames and gives as an output the classification based on the trained labels.

For the training methodologies for both models, Kaggle was used as a standard environment. This allows for an easier process for importing and using the dataset since there was no possibility of the environments of each model colliding with each other.

For the final framework that combines both models some of the code was reused. The libraries needed were torch [11], librosa [12], transformers [13], AV [17], moviepy [19], torchaudio [20][21]. 

For the framework shown in (Fig. 1), the complete pipeline starts with a function that separate the visual and auditory parts from the input videos, then the functions to prepare the videos are re-utilized, the videos gets prepossessed, then the sentiment analysis function gets the emotion string from the label int, The probabilities for each emotion are found using the probability function. On the other hand the audio gets prepossessed and the probabilities for each emotion are calculated as well ,and the pre-trained models are loaded and configured appropriately.

Finally, the main part of our testing framework gets all the file paths from the data folder that end in mp4 and separates them into video-only and audio-only. Each of these files is fed to its respective model and the prediction with the highest probability is taken from the resulting output and kept as that model's prediction. For the combined prediction, all the probabilities for each emotion outputted by the models are taken to compute the mean value. The emotion with the highest resulting probability is then considered the final prediction outputted by the work of both models.

\begin{figure}[h!]
    \centering
    \includegraphics[width=0.85\textwidth]{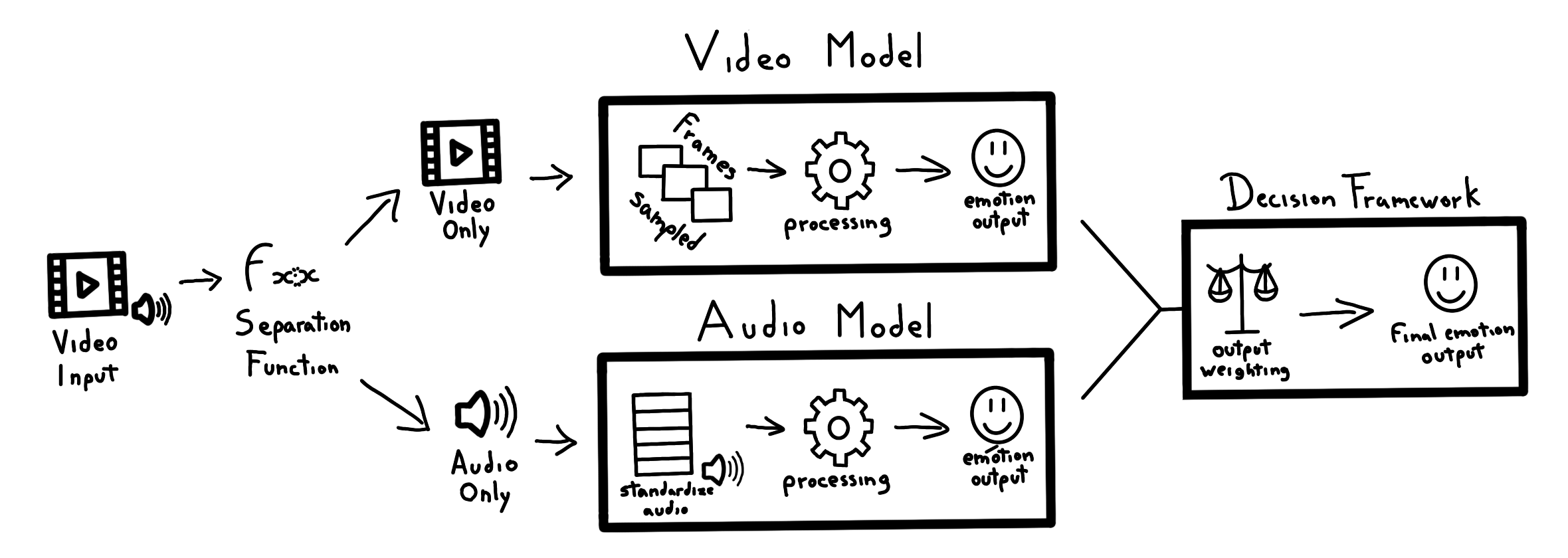}
    \caption{The entire process of our module}
    \label{fig:modelgraph}
\end{figure}

\section{ Results}

\subsection{ Audio Model Training}

To train the audio model, the following main libraries were imported: torch [11], librosa [12], transformers [13], datasets [14], sklearn [15], and tensorflow [16]. 

For training the audio model the following steps were taken: an EDA was performed to understand the data, the model and its processor were loaded, the model configuration was modified to work with our six labels,  two functions were created an used to extract the features and labels from the data into a Dataset object, the dataset was split into train and test with a test size of 0.2, a data collator and preprocessing function were created to adapt the audios to our input data, and finally a compute metric function was created to get the metrics for every 500 steps.

Finally, we define the parameters of our model trainer. We select the evaluation strategy as steps so that we can continue to monitor how our model performs as more data keeps getting fed into it. We trained the model on 6 epochs because this is the maximum number of epochs that we could perform before it overloaded the memory that we had available. Finally, we define the saving steps parameter to 500 to log how our model improves every 500 steps. After roughly 1 hour and 15 minutes of training on our data, the model reached around 72.59\% accuracy on our test set. You can see the progression and result of this training in the Kaggle notebook Audio Sentiment Analysis Model Training [23] at table(2):

\begin{table}[h!]
    \centering
    \caption{Results from training the audio model}
    \begin{tabular}{lcccccc}
        \toprule
        Step & Training Loss & Validation Loss & Accuracy & F1 & Precision & Recall \\
        \colrule
        500 & 1.487500 & 1.310749 & 0.539960 & 0.506796 & 0.585382 & 0.551210 \\
        1000 & 1.036900 & 1.171982 & 0.620551 & 0.608111 & 0.651514 & 0.626476 \\
        1500 & 0.769400 & 1.041908 & 0.689053 & 0.684392 & 0.699751 & 0.695532 \\
        2000 & 0.541500 & 0.969430 & 0.722633 & 0.718879 & 0.725272 & 0.727914 \\
        \botrule
    \end{tabular}
\end{table}
\subsection{ Video Model Training}

For the vivit-b-16x2-kinetics400 video model, a similar code was implemented. First, the following main libraries were imported: torch [11], transformers [13], sklearn [15], AV [17], tdqm [18]. 

For training the video model the following steps were taken: an EDA was done to understand the data, the calm and surprised emotions were removed from the dataset to maintain the same 6 emotions, two functions were created to handle the labels preprocessing, two functions are modified from the vivit transformers documentation to resize the frames to 224 by 224 in order to be compatible with the model, another function was created to select the frames from the video, the dataset class was also created with the necessary parameters, the dataset was divided into training and testing with a testing size of 20\%, data loaders were also created to populate our dataset objects with a batch size of 1 to not overload the Kaggle GPUs, the model was loaded and configured to handle only 6 variables, the training loop was defined so that it was trained for 10 epochs, after each epoch the model is evaluated with the average training and validation loss.

After roughly 7 hours of training for 10 epochs, the model achieved a training loss of 0.1460 and a validation loss of 0.4049. You can find more information about the coding process in the Kaggle notebook Video Sentiment Analysis Model Training [24].

\subsection{ Framework Scores}

After testing the framework (Fig. 2), it can be observed that the video and audio models have similar accuracy scores, although the audio model tends to come on top. As for the combined approach, the results are higher than those of both models.

However, after retraining both models again we managed to end up with a video model that has higher accuracy on the test set (around 88\%) and an audio model that has a similar accuracy on the test set (around 59\%). This led to some problems because when we ran my initial testing framework that worked with the aforementioned probability averaging method, the results were different. Both versions were run on our test sample of 105 files from the RAVDESS dataset. 

Since the combined approach of this method ended up being lower than using just the video model for V2, we decided to try to come up with some other potential decision methods for the framework. The following results are going to include the metrics for both the initial trained models that had similar accuracies, which we are going to call V1, and the newly trained models in which the video accuracy is higher, which we are going to call V2.

The Weighted Average method (Fig. 3) consists of the averaging of the probabilities but this time scaling it according to the respective accuracies of each model. 

The Confidence Level Threshold on the Video Model approach (Fig. 4) involves giving more priority to the video method since it is the most accurate. This is done by taking the video prediction as the final answer if it has a confidence level of more than 0.7 while doing the average probability method if not. 

The Dynamic Weighting Based on Confidence method (Fig. 5) involves using a dynamic weighting method based on the confidence of the predictions. The confidence value for each prediction is calculated by dividing the probability of the predicted label by the total of all predicted probabilities. When the predictions of our models are combined, the weights are assigned based on the inverse of their total confidence scores. 

The Rule-Based Logic approach (Fig. 6) consists of implementing a special rule logic to get the final prediction. If both models agree on the emotion and both are over the desired confidence level of 0.5, the agreed emotion is returned as the final prediction. Otherwise, the emotion is selected based on which model has the highest confidence level. 

You can see the results of these tests in the Kaggle notebooks Multimodal Sentiment Analysis Test Framework V1 [25], Multimodal Sentiment Analysis Test Framework V2 [26], and the graphs below:
\begin{figure}[h!]
    \centering
    \begin{minipage}{0.3\textwidth}
        \centering
        \includegraphics[width=\textwidth]{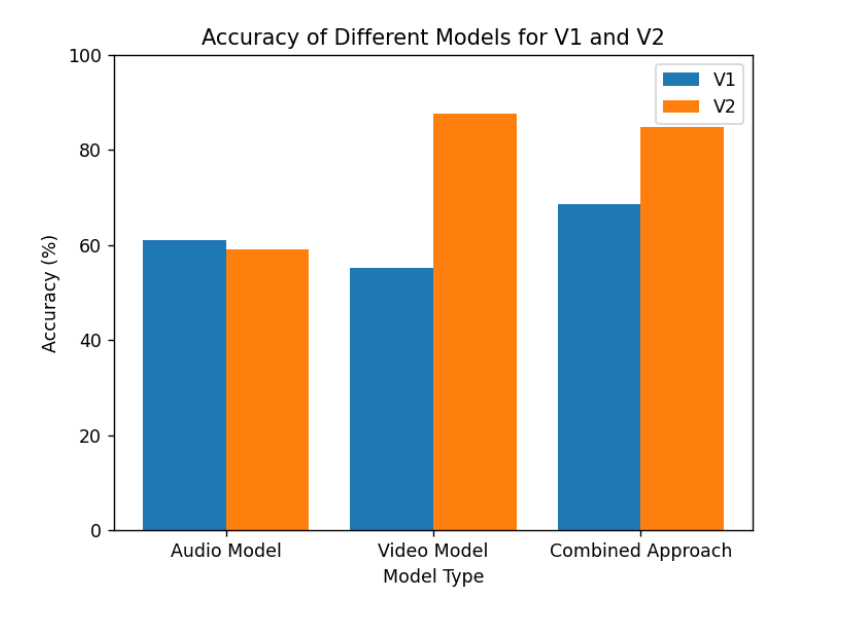}
        \caption{Accuracy of Different Models Using Averaging Method}
    \end{minipage}%
    \hfill
    \begin{minipage}{0.3\textwidth}
        \centering
        \includegraphics[width=\textwidth]{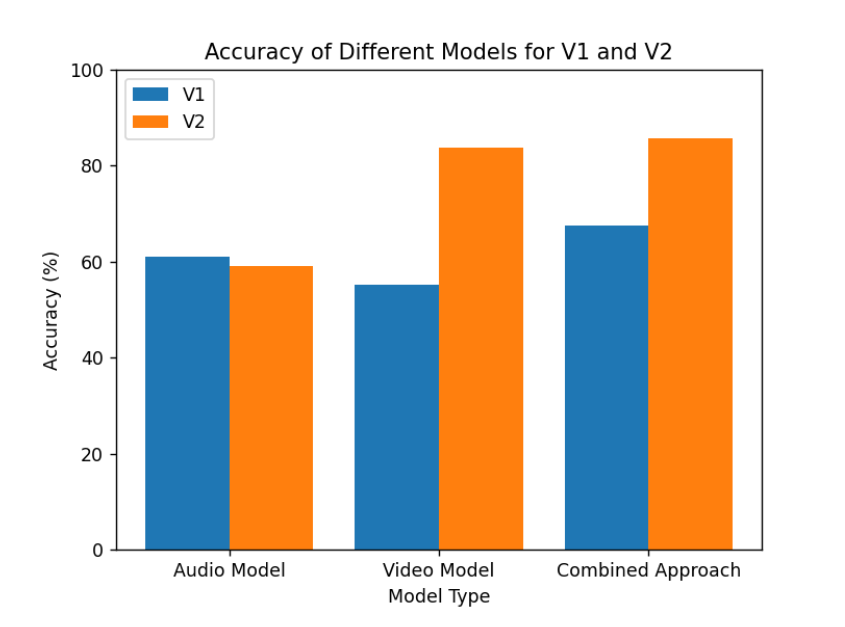}
        \caption{Accuracy of Different Models Using Weighted Average Method}
    \end{minipage}
    \hfill
    \begin{minipage}{0.3\textwidth}
        \centering
        \includegraphics[width=\textwidth]{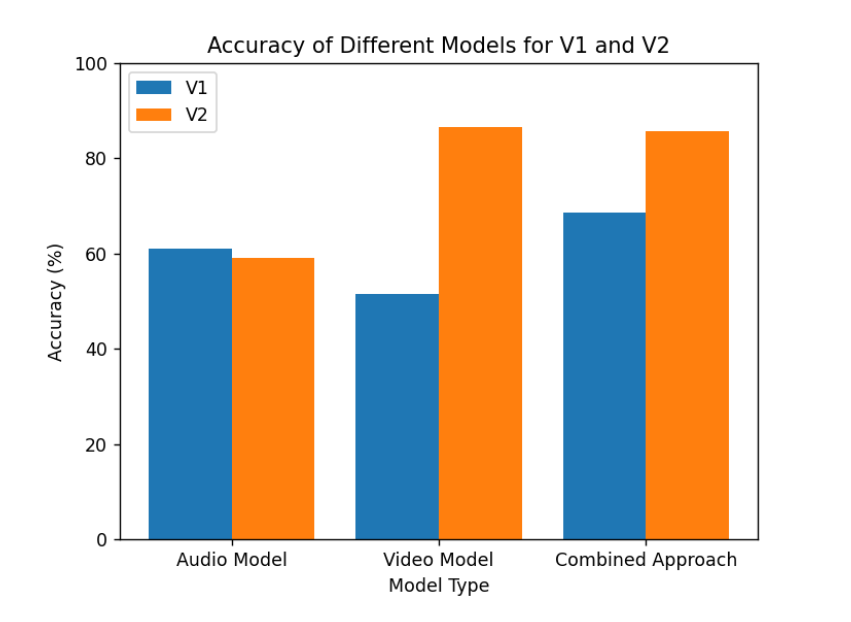}
        \caption{Accuracy of Different Models Using Confidence Level Threshold on the Video Model Method}
    \end{minipage}
    \vspace{0.3cm}
    \begin{minipage}{0.3\textwidth}
        \centering
        \includegraphics[width=\textwidth]{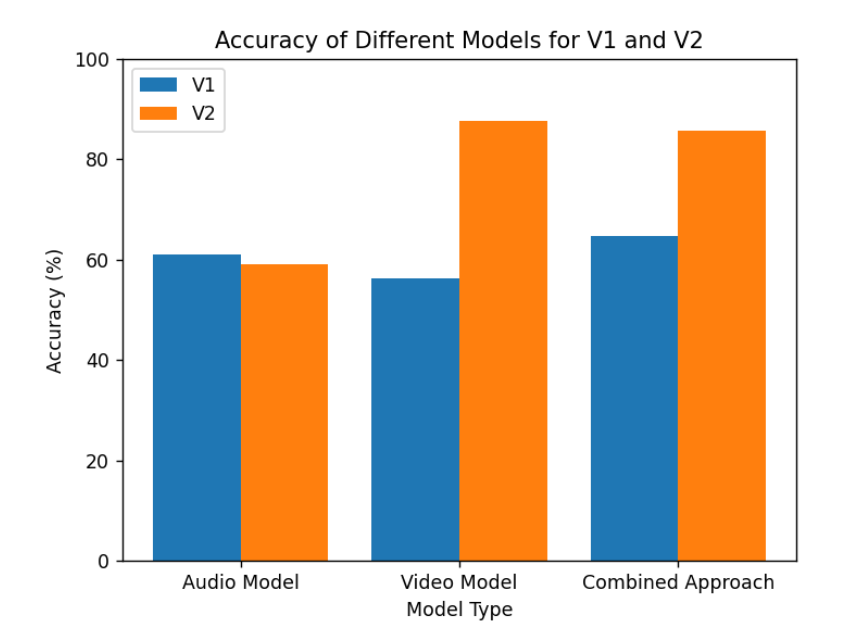}
        \caption{Accuracy of Different Models Using Dynamic Weighting Based on Confidence Method}
    \end{minipage}
    \hfill
    \begin{minipage}{0.3\textwidth}
        \centering
        \includegraphics[width=\textwidth]{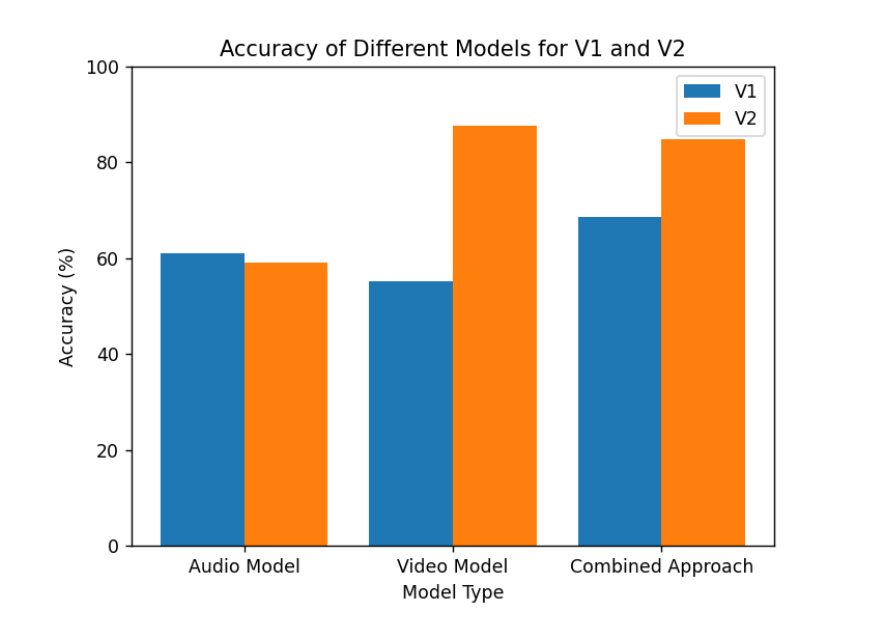}
        \caption{Accuracy of Different Models Using Rule-Based Logic Method}
    \end{minipage}
\end{figure}

\section{Discussion and Conclusion}

After testing, we can observe that the models which have similar results in accuracy are improved when used together in our 5 different decision-based frameworks. However, it seems that the decision frameworks that offer the better results are the Averaging Method and the Rule-Based Logic method. This is because when both models have similar accuracy the averaging of the probabilities tends to have more weight than the other methods. Something similar happens with our implemented Rule-Based Logic since it is based on the confidence intervals of both models and takes into account whether the models strongly agree with each other. On the other hand, when one of the models is more accurate than the other one by a lot, we can see that it is harder to get better results. The Averaging Method got slightly worse results since the less accurate audio model got the same level of priority as the video model. A similar situation happened with the Dynamic Weighting Based on Confidence and Rule-Based Logic methods. The Confidence Level Threshold method tended to be around the same value as the video model due to the prioritization of the video model’s outputs if they were above the 0.7 confidence level. And the Weighted Average Method was the one that tended to give better results since at minimum it got the same score as the video model and sometimes it surpassed it. In the end, the combined approach of using video and audio inputs gave acceptable results on the emotion recognition tasks. However, further testing would be required to see the consistency of these results.

As for the limitations. The video dataset focused on people from a single country in a controlled room where there were not any visual obstacles or changes in the illumination. This could lead to a generalization problem if the trained model were to be done in the wild. By being based on a dataset from a single culture, the model might be biased to certain types of ways of expressing emotions. This could be fixed by gathering more data from people who were raised in different countries and environments. Also, by doing it in a controlled environment the model might not be robust to handle potential noise. This noise could be other people walking in the background or different types of illumination quality. This could also be fixed by making sure that our new data entries are collected in the wild and cover a wide margin of different scenarios. Another potential fix could be to add a facial recognition model that detects the most relevant face in the frame and crops it to feed it to the model. The training of both models was done in the free version of Kaggle. This offered a lot of advantages in terms of handling the CUDA environment for the datasets and models. But it also came with constant memory overloads due to the great but limited memory capacity that the GPUs on Kaggle had. If these models were to be trained on better hardware, they might achieve a better performance.

Some future work that could be done is to try to add a third model that gets the emotion recognition based on doing Natural Language Processing techniques on the transcribed text of the audio. The experiment could try to measure how the three models act together and also try to group them up into different pairs (audio-text, audio-video, and text-video) to see which conjoined architecture offers the best performance. Another implementation would be to try to use this model architecture inside a robot so that it could act as a therapy companion for people with mental health issues. This could be done by reformatting our framework to take live video feeds and know the emotions in real time of the user. This emotion could then be passed into a generative model for the robot to take into account alongside his current conversation with the user to give a response that is more mindful of the user’s state of mind. However, it is also worth noting that the AI Act was passed by the EU. This act deemed emotion recognition systems to be of high risk due to their poor generalizability and their potential for discrimination as per page 41 of the Artificial Intelligence Act of the European Parliament (2024) [22]. Therefore some legal advice should be taken to make sure that everything is done ethically and according to the law. Although, the use that was initially thought for this project, the implementation of a therapeutic companion robot to treat mental health issues, seems to be under the non-prohibited uses of AI for emotion recognition as per page 41 of the Artificial Intelligence Act of the European Parliament (2024).

In conclusion, emotion recognition that focuses on video and audio inputs showed acceptable results under limited conditions, but further research with more resources could help shed light on the usability of this approach.

\section*{Acknowledgements}
I want to thank my supervisor and the teachers at IE for all the help and knowledge they gave me over the past few years. This paper is dedicated to my family for all the love, support, and encouragement that they gave me and have allowed me to do what I love. And lastly, I would also like to thank my friends for their encouragement throughout this project.

\clearpage

\end{document}